\documentclass[twoside]{procinasan}

\input{procinasan.def}

\begin{document}

\pagebreak

\thispagestyle{titlehead}

\setcounter{footnote}{0}
\setcounter{equation}{0}
\setcounter{section}{0}
\setcounter{figure}{0}
\setcounter{table}{0}

\selectlanguage{english}

\addcontentsline{toce}{subsection}{{\em Potravnov I.S., Ryabchikova T.A.\/}
On the surface distribution of chromium in Ap star HD~152564}


\titlen{On the surface distribution of chromium in Ap star HD~152564}
{Potravnov I.S.$^{1*}$, Ryabchikova T.A.$^1$}
{$^1$Institute of Astronomy of the RAS, Moscow, Russia\\
*e-mail: ilya.astro@gmail.com}

\abstract{The spectroscopic and photometric variability of magnetic Ap/Bp stars is caused by the surface inhomogeneities of their chemical composition (spots) that affect the structure of the atmosphere and the emergent flux. Interpretation and modeling of the variability of Ap/Bp stars requires careful consideration of the surface distribution of the principal elements which contribute to opacity in a given temperature range. In this paper we present the results of a study of the horizontal distribution of chromium in the atmosphere of Ap star HD~152564 reconstructed with the Doppler Imaging technique. We reveal that the region of increased chromium abundance forms a ring perpendicular to the rotational equator. The passage of these regions across the visible hemisphere of the star should contributes to the observed light variability. Comparison with the previously obtained results shows that the character of the horizontal distribution of chromium in the atmosphere of HD~152564 is different from distribution of other investigated elements, which in turn also possess a significant diversity.

\medskip

\noindent 

\medskip


\medskip

\noindent {}

\selectlanguage{english}

\baselineskip 12pt
\section*{1. Introduction}

Chemically peculiar Ap/Bp stars are known to host the large scale magnetic fields up to $B\sim10^4$ G strength and also possess anomalies of surface chemical composition.
These anomalies are ussualy manifest as deficit of light elements as well as overabundance of silicon, iron peak and rare earth elements up to $\sim$5~dex in logarithmic abundance scale relative to the solar composition \cite{2007AstBu..62...62R}. The elemental distribution in Ap/Bp atmospheres is characterised by significant vertical and horizontal abundance gradients which are formed by the selective atomic diffusion \cite{1970ApJ...160..641M,2015ads..book.....M}. The photometric, spectroscopic and magnetic variability of Ap/Bp stars is modulated with the period of axial rotation is caused by the passage on the line of sight the chemically-inhomogeneous (spotty) stellar surface \cite{1950MNRAS.110..395S}.\\
Local elemental overabundances modify the thermal structure of the atmosphere by the line blanketing and affect the emergent flux \cite{2007A&A...469.1083K}. In the temperature domain of Ap/Bp stars the silicon and some iron peak elements like iron and chromium with significant number of transitions observed in the visual range are the principal contributors to opacity. Thus, the inhomogeneous surface distribution of these elements contributes significantly to the spectroscopic and photometric variability and should be taken into account during the modelling of the light curves of Ap/Bp stars.\\ 
The silicon Ap star HD~152564 (=MX TrA) possesses the notable variability of spectral lines profiles as well as the quasi-sinusoidal light variations with $\Delta I\approx0.03$~m amplitude and period $P=2.1639$~d. The abundance analysis revealed the significant overabundance of Si, Fe, Cr in its atmosphere \cite{2024MNRAS.52710376P}. The Doppler Imaging (DI) of HD~152564 was also performed by Potravnov et al. \cite{2024MNRAS.52710376P} in the lines of five elements (He, Mg, Si, Fe, O) and revealed their strongly inhomogeneous distribution with abundance contrasts up to $\sim2$ dex. Moreover, the horizontal distribution of the examined elements turned out to be significantly different. In the present work, we supplement the previous results by investigation of the horizontal distribution of chromium in HD~152564 atmosphere.             

\section*{2. Doppler Imaging}

We used the Doppler Imaging (DI) technique realized in INVERS11 program package \cite{2002A&A...381..736P} for the inversion of the surface distribution of chromium in HD~152564. INVERS11 allows to recover the surface distribution of given element with the fitting of observational line profiles obtained at different rotational phases. The spectral synthesis of local line profiles is performed with the adopted model atmosphere and chemical composition within the elements of equidimensional grid covering the stellar surface. The abundance of the examined element is the free parameter. The disk integration is performed taking into account the Doppler shifts. Because DI is the classical ill-posed problem, the inversion is solved with the Tichonov's regularization of $\chi^2$ minimisation of residuals between synthetic line profiles and observational ones. In the inversion procedure the value of the regularization parameter that provides a minimum value of $\chi^2$, comparable to the S/N of the original data is usually employs. INVERS11 allows to recover the surface distribution of several elements simultaneously, which allows to employ blended lines to analysis.       

\begin{table}[h]
\center
\caption{Parameters of HD~152564}
\label{tab1}
\begin{tabular}{lc}
\hline
Parameter & Value \\
\hline
\hline
$T_{eff}$ & 11950$\pm$200 K \\
\lgg  & 3.6$\pm$0.2 dex \\
$V_\zeta$  & 0.0~km s$^{-1}$ \\
$V_{RT}$  & 0.0$\pm$1.5~km s$^{-1}$ \\
\vs    & 69$\pm$2~km s$^{-1}$ \\
$i$    & 51$^{\circ}$\\

\hline
\end{tabular}
\end{table}

For the mapping of surface distribution of chromium in HD~152564 we used the spectroscopic time series obtained in 2019--2021 with the 10-m. SALT telescope in South Africa equipped with the HRS echelle spectrograph. The details on the observations and data reduction see in \cite{2024MNRAS.52710376P}. The spectra were obtained in 10 equidistant rotational phases in the medium resolution mode of HRS ($R\approx38000$). The data were phased using the ephemeris $JD=2458647.7774 + 2.1639d\cdot E$ where the initial epochue corresponds to the maximum light. The list of stellar parameters for the model atmosphere calculations and DI is presented in Tab.~\ref{tab1}. Also, the mean abundances in HD~152564 determined in \cite{2024MNRAS.52710376P} were used for model atmosphere calculations and as an initial approximation in the DI procedure. Given the relatively rapid (for Ap stars) axial rotation of HD~152564 (\vs=69 km/s), it is difficult to find lines in its spectrum suitable for DI. The CrII lines 4558.62\AA\,, 4616.63\AA\, and 4634.07\AA\, which have reliable atomic parameters, were chosen to study the distribution of chromium on the surface of HD~152564. The atomic parameters of the examined CrII lines and blending them FeII, SiIII, and TiII lines were obtained from the VALD3 atomic database \cite{1995A&AS..112..525P,2015PhyS...90e4005R} and are presented in Table \ref{tab2}. The model of the HD~152564 atmosphere calculated in the \textsc{LLmodels} software package \cite{2004A&A...428..993S} and accounting for the individual chemical composition of HD~152564 was used for the spectral synthesis. \\  

\begin{table}[h]
\center
\caption{Linelist for DI}
\label{tab2}
\tabcolsep=3pt
\begin{tabular}{lcccc}
\hline
Element & Line, \AA\, & $E_{low}$, eV & $\log gf$ & $\log \Gamma_4$ \\
\hline
\hline

SiIII & 4557.2472 & 28.3554 & -0.070 &  0.000\\
CrII & 4558.2796 &  8.3557 & -1.272 & -5.730\\
CrII & 4558.6241 &  4.0734 & -0.430 & -6.540\\ 
CrII & 4558.7571 &  4.0735 & -2.530 & -6.540\\
FeII & 4559.5193 &  7.8042 & -2.447 & -5.830\\
...\\
FeII & 4615.3230 & 7.6951 & -3.112 & -5.850\\
CrII & 4616.2582 & 5.6699 & -2.346 & -6.480\\
CrII & 4616.6290 & 4.0722 & -1.420 & -6.560\\
FeII & 4618.0591 & 7.6951 & -3.313 & -5.850\\
...\\
CrII & 4634.0700 & 4.0722 & -1.050 &  0.000\\
FeII & 4634.6058 & 2.5827 & -5.533 & -6.540\\
FeII & 4635.3169 & 5.9561 & -1.476 & -6.580\\
TiII & 4636.3219 & 1.1649 & -3.230 & -6.500\\
\hline
\end{tabular} 
\end{table} 

\begin{figure}[h]
\begin{center}
\includegraphics[width=0.7\textwidth]{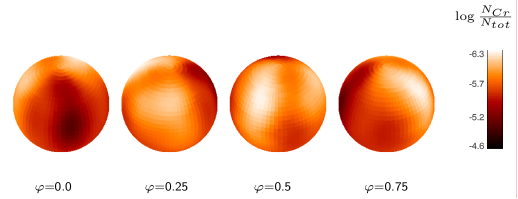}
\caption{Surface distribution of chromium in HD~152564 atmosphere from CrII 4558.62\AA\,, 4616.63\AA\, and 4634.07\AA\, lines. The abundance scale in $\log A_{Cr}=\log(N_{Cr}/N_{tot})$ units is shown at the right.}
\label{fig1}
\end{center}
\end{figure} 

Because the CrII lines used for DI are blended by FeII, we performed simultaneous reconstruction of surface distribution of chromium and iron. The distribution of iron appeared to be identical to those obtained in \cite{2024MNRAS.52710376P} from unblended FeII 4731.45\AA\, line. This confirms that in our analysis the contribution of chromium and iron to the blend variability is treated correctly. The results of chromium mapping over the HD~152564 surface are shown in Fig.~\ref{fig1}, which presents maps in spherical projection for the four rotational phases $\phi=0.0-0.75$. One can see that the surface distribution of chromium in HD~152564 atmosphere is highly inhomogeneous with abundance contrasts up to $\sim1.7$ dex. The regions of Cr overabundance form the relatively narrow ($\Delta l\approx20^\circ$ in longitude) meridional ring which transits across the line of sight at phases $\phi\approx0.0$ and 0.5. The ring has a clumpy structure with the most contrast spot with $\log A_{Cr}=\log(N_{Cr}/N_{tot})\approx-4.6$~dex located at the intersection with the rotational equator near phase $\phi=0.0$. The chromium abundance in this spot is almost two orders of magnitude higher than those in solar atmosphere ($\log A^{\odot}_{Cr}=-6.38$~dex). The other regions possess Cr abundance close to the solar one. An example of fitting of observational CrII lines profiles at different rotational phases is shown in Fig.~\ref{fig2}. One can see the good match between observed profiles and synthesis. The profiles of the CrII line 4558.62\AA\, showing additional absorption in comparison with the synthetic spectrum at phases $\phi=0.44$ and 0.98 are the exception. At the same time, there is no additional absorption in the other two CrII lines we used. We have no clear explanation for this circumstance. We constructed a map using the two CrII lines 4616.63\AA\, and 4634.07\AA\, excluding the CrII line 4558.62\AA\ (although it is the least blended amoung the three examined lines) for verification. Results are shown in Fig.~\ref{fig3}. The overall character of the distribution remained the same, but the contrast of the dominant region at the phase $\phi=0.0$ has changed. The maximum Cr abundance decreasing by $\sim0.5$ dex, causing the smoothing the gradient in the spot and makes it visually more homogeneous and diffuse. The shape and abundance of the other regions remained almost unchanged. It should be noted that the resulting uncertainties in profiles fitting considering the CrII 4558.62\AA\ line and excluding it are practically identical. Currently, it is not possible to make an unambiguous choice between the two presented Cr abundance scales. Perhaps further implementing of additional data, e.g., light curve modeling, will allow to solve this issue.

\begin{figure}[t]
\begin{center}
\includegraphics[width=0.3\textwidth, angle=180]{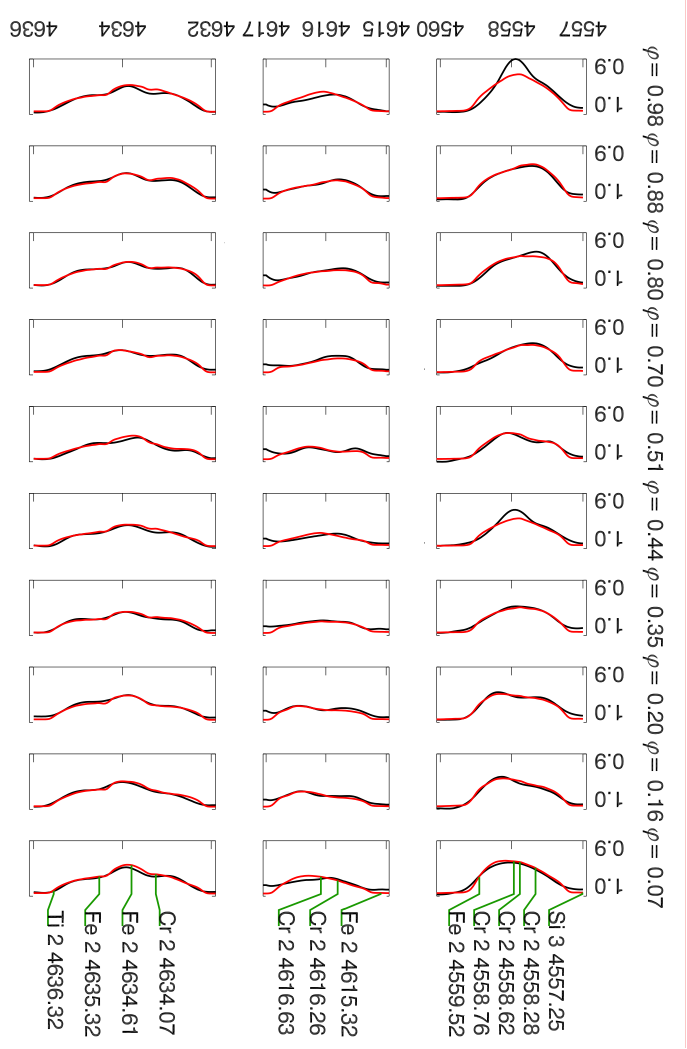}
\caption{CrII line profiles in the HD~152564 spectrum at different rotational phases (black line) and their approximation by the synthetic spectrum (red line). The phases are labeled left from the figure.}
\label{fig2}
\end{center}
\end{figure}

\begin{figure}[h]
\begin{center}
\includegraphics[width=0.7\textwidth, angle=0]{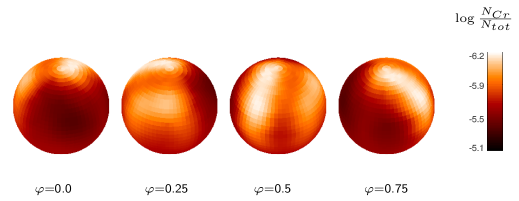}
\caption{The same as Fig.~\ref{fig1}, but from CrII 4616.63\AA\, and 4634.07\AA\, lines only. When comparing with Fig.~\ref{fig1}, note the changed abundance scale and dynamical range of the colorbar.}
\label{fig3}
\end{center}
\end{figure}

\section*{3. Discussion and conclusions}

The DI revealed a highly inhomogeneous distribution of chromium over HD~152564 surface with significant abundance contrasts. Usage of different sets of lines results in $\sim0.5$ dex difference in abundance scale, but almost identical character of distribution. The region of increased Cr abundance appears as an inhomogeneous ring along a large circle perpendicular to the equator of rotation. This character of Cr surface distribution differs from the previously studied distributions of other elements: chains of near-equatorial spots (Si, O) and circumpolar ring-like structures (He, Fe, Mg). The region of maximum Cr abundance does not coincide with any of the elemental spots previously mapped in the atmosphere of HD~152564, and therefore its distinct contribution to the photometric variability is expected. This issue will be quantitatively examined in subsequent work.\\
It is interesting to note that according to predictions of modern three-dimensional diffusion models \cite{2016MNRAS.457...74S}, narrow ring-like structures should appear along the magnetic equator. Moreover, the same distribution character is expected for all elements. In fact, the observed surface elemental distributions in Ap/Bp stars are characterised by a great diversity, which does not agree with the model predictions (see the discussion in \cite{2018MNRAS.474.2787K}). With our multi-element mapping, the case of HD~152564 clearly shows that even in the atmosphere of one star, the horizontal distributions of elements with similar absorption spectra, such as Fe and Cr, can be significantly different. Since the characteristics of the magnetic field of HD~152564 are unknown, based on the observed chromium distribution and the predictions of diffusion models, we could suggest a large dipole tilt ($\beta\approx90^\circ$) and the location of the 'chromium ring'\ in the plane of the magnetic equator, perpendicular to the equator of rotation. However, this assumption is inconsistent with the iron concentration in the north polar region. Thus, we can conclude that a promising task is the Doppler-Zeeman mapping of HD~152564, which would allow to reconstruct the surface geometry of the magnetic field and study associated anomalies of chemical composition. These results would be important for further testing the modern diffusion models.

\bibliographystyle{inasan}
\small
\bibliography{ref}
\normalsize

\end{document}